\documentclass[pre,amsfonts,amssymb,amsmath,floatfix,nofootinbib,twocolumn,showpacs,superscriptaddress]{revtex4}
\usepackage{amsmath,psfrag,graphicx,color}
\begin{document}
\title{Traffic of cytoskeletal motors with disordered attachment rates}
\author{H. Grzeschik}
\thanks{Deceased.}
\affiliation{Fachrichtung Theoretische Physik, Universit\"at des Saarlandes, 66041 Saarbr\"ucken, Germany.}
\author{R. J. Harris}
\email{rosemary.harris@qmul.ac.uk}
\affiliation{Fachrichtung Theoretische Physik, Universit\"at des Saarlandes, 66041 Saarbr\"ucken, Germany.}
\affiliation{School of Mathematical Sciences, Queen Mary University of London, Mile End Road, London, E1 4NS, United Kingdom.}
\author{L. Santen} 
\email{santen@lusi.uni-sb.de}
\affiliation{Fachrichtung Theoretische Physik, Universit\"at des Saarlandes, 66041 Saarbr\"ucken, Germany.}

\date{\today}

\begin{abstract}

Motivated by experimental results on the interplay between molecular
motors and tau proteins, we extend lattice-based models of
intracellular transport to include a second species of particle which
locally influences the motor-filament attachment rate.  We consider
various exactly solvable limits of a stochastic multi-particle model
before focusing on the low-motor-density regime.  Here, an approximate
treatment based on the random walk behaviour of single motors gives
good quantitative agreement with simulation results for the
tau-dependence of the motor current.  Finally, we discuss the possible
physiological implications of our results.
\end{abstract}


\maketitle
\section{Introduction}
\label{s:intro}

Active transport on a microscopic scale is one of the most important 
features of living organisms~\cite{Alberts02}. Transport of this kind is responsible for 
the functionality of eucaryotic cells and ultimately, via contraction of 
muscle cells, for the motion of organisms. Unsurprisingly, perturbations 
of the intracellular transport can lead to pathological conditions; for
example, the role of molecular motors in sensory defects is discussed
in~\cite{Avraham03}. Another example is Alzheimer's disease in which the breakdown of axonal 
transport, taking place before plaque formation can be observed,
has been reported for transgenic mice~\cite{Stokin05}.  In particular, there is
experimental evidence that an over-expression of the linker-protein
tau may reduce the tubulin affinity of kinesin motors~\cite{Trinczek99,Seitz02}.   Therefore the investigation of 
the basic mechanisms of intracellular transport is not merely of theoretical 
interest.  

Studies of molecular motors discuss, e.g., the necessary prerequisites for the 
directed stochastic motion of single motors or the cooperative dynamics of 
many motor proteins attached to a single
cargo~\cite{Julicher97,Mallik04b,Singh05,Klumpp05c}. From a theoretical point of
view these cases can be considered as single self-driven particles. Although 
the characterization of the individual motions is of high relevance for many
biological and biophysical systems, often intracellular transport is carried 
out by several self-driven particles. Systems of this kind show generic 
many-particle behaviour~\cite{Nishinari05}. Focusing on the many-particle features of 
intracellular transport, it is sufficient to consider an effective motion 
for the molecular motors on the filaments. Molecular motors move 
generically unidirectionally and stepwise~\cite{Schliwa03}. The size
of the steps is, in general, load dependent and given 
by multiples of the structural unit length of the filament. Another important 
feature of the motor dynamics is its stochastic nature, i.e., binding and 
unbinding as well as the movements on the filament are random events which are described 
by rates.  

These properties of the dynamics suggest that intracellular transport
can be described by models that are in close analogy to the asymmetric
exclusion process (ASEP)---a one-dimensional lattice model for
interacting self-driven particles~\cite{Derrida98}.  Compared to the
ASEP, one has to account for the finite ``run length'' of molecular
motors, i.e., the absence of particle conservation on the track.
Various models of this kind have been proposed, see
e.g.~\cite{Lipowsky01, Parmeggiani03}, which show particularly
interesting many-particle effects on open lattices. The absorption and
desorption of particles leads to the formation of localized high and
low density domains.  This modelling approach is very flexible and can
be easily generalized, for example, in order to explain the results of
\emph{in vitro} experiments~\cite{Nishinari05}.  For recent reviews
treating statistical mechanics descriptions of intracellular traffic,
see e.g.,~\cite{Hinsch07,Chowdhury07e,Klumpp07}.

In the present work we discuss the influence of a second type of
particle which alters the attachment rates of the self-driven particles
in different one-dimensional environments. The model under
investigation aims to reproduce the generic features of the kinesin
dynamics on microtubules for different concentrations of tau.  In
addition to shedding light on this specific biological situation our
study contributes to the understanding of disorder in driven
many-particle systems.

The rest of the paper is organized as follows.  In Sec.~\ref{s:sim}
we review the lattice-based model for intracellular transport
introduced by Klumpp and Lipowsky~\cite{Klumpp03} and discuss
simulation results obtained by generalizing it to include the effect
of tau.  Then, to gain further insight, we introduce in
Sec.~\ref{s:model} a somewhat simplified model which retains the
key features of the problem but is more amenable to analytical
treatment; we also discuss some limits in which this model is exactly
solvable.  In Sec.~\ref{s:tauvary} we address the biologically
important question of how the mean current and its fluctuations depend
on the concentration of tau.  A random walk treatment for single
motors leads to predictions for the low-motor-density regime of the
multi-particle model which are in good agreement with simulation.
Finally, in Sec.~\ref{s:diss}, we conclude by discussing the possible
physiological relevance of our results and potential generalizations to
more realistic models.

\section{From biology to simulation}
\label{s:sim}

In this section we outline how the complex biological system of
interest can be represented by a lattice-based model amenable to
simulation.  Specifically, we build on the work of Klumpp and
Lipowsky~\cite{Klumpp03}, generalizing their model to include tau
proteins and presenting specimen simulation results.  This is in the
spirit of a now well-established approach to complex systems whereby
one considers simple models in order to help understand phenomena and
mechanisms which may also occur in more realistic situations.  Indeed,
we will pursue this approach further in the following section in which
we discuss an even more simplified model allowing for analytical
treatment.

\subsection{Model of Klumpp and Lipowsky}

As discussed in the introduction, motor proteins carry out directed
walks (i.e., active transport) on the filaments of the cytoskeleton.
In the context of neuronal transport, we are particularly interested
in the motion of kinesin along axon microtubules (composed of
repeating tubulin subunits)~\cite{Gunawardena04}.  The stepwise
structure of such movement naturally suggests using a one-dimensional
(1D) lattice model with lattice-spacing equal to the step size~\cite{Schliwa03}.
We
assume that the size of the kinesin molecules is comparable to this
unit lattice spacing; the mutual exclusion of motors is then imposed
by preventing double occupancy of any one site.  An obvious starting
point for modelling the dynamics is the prototypical asymmetric simple
exclusion process in which particles hop to vacant nearest neighbours
with some preferred direction (for a review of this exactly-solvable
model, see~\cite{Derrida98c}).  In fact ``backward'' steps for
kinesin are extremely rare so it suffices to consider the totally asymmetric
version in which particles can only hop in one direction.

Since kinesin motors typically detach from the microtubule after a
certain number of steps (``run length'') and diffuse in the
surrounding medium before reattaching, one needs to extend the
standard ASEP picture.  A simple choice is to impose the same lattice
structure and exclusion rules off the filament but allow the molecules
to move symmetrically there.  The association onto and dissociation
from the track is governed by explicit rates (typically the
dissociation rate is comparatively small so that a single motor can
make many steps before detaching).  

Klumpp and Lipowsky embed this lattice model in a cylindrical geometry with directed motion
along a 1D track located at the symmetry axis of the cylinder and
undirected motion (diffusion) elsewhere.  Of course, this is a
considerable simplification compared to a real nerve cell which
contains a bundle of parallel microtubules.  However, it is expected
to reproduce qualitatively the main features of unidirectional
transport (and could readily be computationally extended to the case of
several parallel tracks).  Another important question
concerns the boundary conditions at the ends of the cylinder.
In~\cite{Klumpp03} both periodic and open boundary conditions are
considered; in the former case, the current-density relation is
obtained and, in the latter, the phase diagram.

\subsection{Effect of adding tau}
\begin{figure}
\begin{center} 
\includegraphics*[width=0.9\columnwidth]{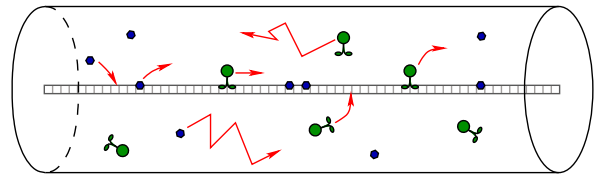}
\caption{(Color online) Graphical representation (not to scale) of the key
  ingredients in the simulation model (after~\cite{Klumpp03} but with the
  addition of tau).  The cylindrical geometry can
  be considered as a simplified representation of a nerve cell where
  the track along the symmetry axis is the axon microtubule.  Our
  simulations are based on a cubic lattice with periodic boundary
  conditions (not shown).  Two-headed kinesin motors (with
  attached cargoes) diffuse in the bulk and undergo directed transport
  along the track.  The tau proteins (represented here by small
  hexagons) also diffuse in the bulk but do
  not move along the track.}
\label{f:geo}
\end{center}
\end{figure}

In this subsection we generalize the above-described model of Klumpp
and Lipowsky to include the effect of tau proteins.  The interested reader is
referred to~\cite{Schweers94,Santarella04} for details of
the structure of tau and its interaction with tubulin.  
Here we merely note that tau molecules ``decorate'' the microtubule
(indeed they are believed to play a role in stabilizing it) but do not
undergo active transport along it.  We
can therefore incorporate them in our model by introducing a second species of
particle which is allowed to diffuse in the bulk,
 and absorb/desorb with certain rates but not to move along the track
itself.  Figure~\ref{f:geo} summarizes the various processes now
  incorporated in the model.
In the present work we restrict ourselves to periodic boundary
conditions.  Although one expects that real nerve cells are better
described by open (or half-open) boundary
conditions~cf.~\cite{Mueller05}, studying the simpler case of periodic
boundary conditions already gives information about bulk effects.  We
further justify our choice by remarking that the finite run lengths
of the molecular motors might be expected to reduce the dependence of
currents on the filament boundary conditions.

To complete our description of the model we must specify the details of the tau interactions.  Firstly we note that tau molecules are relatively small
compared to the motor-cargo complex.
Single motor \emph{in vitro}
experiments~\cite{Seitz02} also indicate that the presence
of tau does not affect the speed of kinesins along microtubules or
their run length (probably because kinesin and tau have different
binding sites~\cite{Santarella04}) but does lead to a significant
reduction in the attachment rate.  

In the present simulation model, we therefore neglect tau-tau and
tau-kinesin exclusion effects (i.e., each unit lattice step can
contain one or more tau particles, regardless of the presence or
absence of a kinesin motor). However, the crucial point is the effect of tau on the kinesin
dynamics.  \emph{In vivo} experiments~\cite{Trinczek99} are
complicated by the fact that many motors can be attached to a single
cargo but also suggest that one effect of tau is a decrease in the
motor attachment rate (especially for kinesin).  
Hence, in the
present model, when one or more tau molecules occupies a site on the
track the rate for absorption of a kinesin is reduced.\footnote{Note that some aspects of our model are similar spirit to the discussion of crowding by another molecular species which already appears in~\cite{Lipowsky06}.  However, the specifics are rather different---in particular, in that work, there is exclusion of motors by the ``obstacles'' and the obstacles can also be actively unbound by processive motors.}

The description above furnishes a model which retains the key features
of the biological system but can be implemented simply in numerical
simulation.  In principle, one can try to match the parameters of the
model (lattice spacing, rates, etc.) to those known from experiments
in order to obtain quantitatively meaningful results.  We carried out
simulations using relative values for the kinesin rates based on data
for the case without tau in~\cite{Lipowsky01}.\footnote{The rates are
equivalent to probabilities per timestep; for comparison with real
data, normalization of the timescale would then be achieved by
associating the length of each timestep with a physical time unit.}
Specifically an unbound kinesin could hop to a neighbouring empty
filament site with rate 0.0083 which was identical to the hop rate
between a pair of empty bulk sites (i.e., the binding probability was
one).  The rate for a forward step on the filament was 0.0099 and the
desorption rate from the filament to each of the four adjacent unbound
lattice sites was $1.7\times10^{-5}$.  The tau molecules were
(arbitrarily) set to diffuse twice as fast as the motors and had
absorption and desorption rates, 0.03 and $2.5\times 10^{-6}$
respectively, which are consistent with data from FRAP (Fluorescence
Recovery After Photobleaching) experiments~\cite{Samsonov04}.  Note
the high affinity of tau for the microtubule which leads to
significant tau coverage on the filament even for low global
densities.

In order to illustrate the possible effect of tau on kinesin binding and transport, we reduced the kinesin adsorption rate by a factor of one hundred in the presence of tau. This value arguably overestimates the impact of tau in real systems.  (In~\cite{Seitz02} it was shown that the  binding rates depend on the ratio between tau and kinesin concentration; for the highest concentration of tau used in that experiment the binding rates were reduced by a factor of five.)  However, we expect a stronger influence of tau in living cells since the typical run length of the molecular motors must be shorter than in \emph{in vitro} assays, due to the bidirectionality of the transport and the dynamics of the filament.

\begin{figure}
\begin{center} 
\includegraphics*[width=1.0\columnwidth]{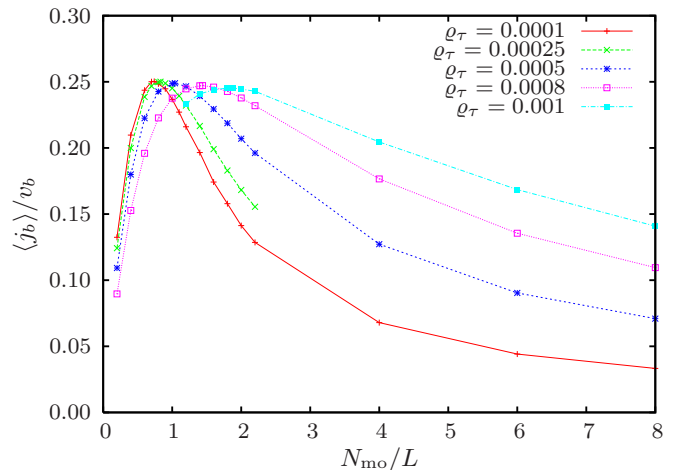}
\caption{(Color online) Average current $\langle j_b \rangle$ along the microtubule
  filament (normalized by the velocity of bound motors $v_b$) as a function
  of kinesin density $N_\text{mo}/L$ for different tau densities
  $\varrho_\tau$. Note that here for consistency with~\cite{Klumpp03}, the \emph{motor} density is measured in number of motors per unit lattice \emph{length} along the axonal cylinder---this can trivially be converted to the number per unit volume using the dimensions of the system (length 1000 lattice sites, radius 25).
Simulation data is averaged over 50
  histories and approximately $2\times 10^6$ time steps.  Lines are
  provided as an aid to the eye.}
\label{f:hj}
\end{center}
\end{figure}

Figure~\ref{f:hj} shows simulation results for the current along the track versus bound kinesin
density for different values of global tau density.
One sees from these results that increasing the density of tau reduces
the height of the maximum and shifts it to higher kinesin densities.
The average current is reduced for low densities of kinesin but
increased for high densities.  Furthermore, the simulation results
cannot be reproduced by a simple mean-field approximation which
indicates the significance of correlation effects.  Indeed explicit
measurements of the kinesin pair-correlation functions show the
existence of long-range correlations.  To explore the tau-induced
effects analytically, in the next section we present and discuss an
even simpler model.

\section{Minimal model}
\label{s:model}

In order to elucidate the mechanisms leading to the features observed
in the simulations described above, we now introduce a simpler model
which can be treated analytically in certain limits.  Specifically,
instead of the cylindrical geometry we consider a two-lane system
(analogous to the ``two-state approximation'' of~\cite{Klumpp03}) with
somewhat simplified effective dynamics for the tau particles.

Note that this is now a stylized ``toy'' model designed to yield
increased understanding but \emph{not} quantitative agreement with
experimental results.  We further remark that simple two-lane models
have previously been used as crude representations of vehicular traffic and
are of more general interest in building up our knowledge of
non-equilibrium statistical physics, see
e.g.,~\cite{Popkov01,Me05c,Pronina06,Hinsch06}.  In the following analysis we
mainly use the language of physics (``lattice sites'', ``particles'',
etc.) whilst keeping in mind the original biological motivation.  The
physiological significance of our results will then be discussed in
more detail in Sec.~\ref{s:diss}.

\subsection{Model definition}
\label{ss:modeldef}

\begin{figure}
\begin{center} 
\includegraphics*[width=0.8\columnwidth]{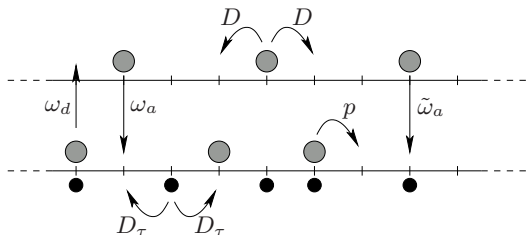}
\caption{Schematic representation of simplified model.  The larger grey circles denote K-particles and the smaller black circles are $\tau$-particles.  Arrows show possible moves with associated rates.  The model has periodic boundary conditions.}
\label{f:model}
\end{center}
\end{figure}

We consider a two-lane lattice gas model with periodic boundary
conditions, as shown schematically in Fig.~\ref{f:model}.  Here the
lower lane represents a microtubule (MT) filament and the upper one
the surroundings.  The model contains two species of particles
representing kinesin motors (K) and tau proteins ($\tau$).   We denote
the occupation number for K-particles on the $i$th site in the lower
or upper lanes by $b_i$ or $u_i$ respectively (where the letters
denote ``bound'' and ``unbound'' in allusion to the biological
context).  The $\tau$-particles are found only in the lower lane (see
below) with occupation number $\tau_i$.   The total number of each
kind of particle is conserved and a hard-core interaction prevents
more than one of each species on a given site, i.e, the site
occupation numbers can take only the values 0 or 1.   The dynamics of
the model (defined in continuous-time) are motivated by the biological
picture, as will be explained below.  

We first make some remarks on the behaviour of the $\tau$-particles.
After the discussion of Sec.~\ref{s:sim} the exclusion condition
may seem very unrealistic.  However, for the kinesin transport the
relevant question is whether a given location on the microtubule
contains \emph{at least one} tau molecule (in our simplified
description the number of such molecules is irrelevant).  It therefore
seems reasonable to reduce the state-space and consider only
``tau-occupied'' and ``tau-empty'' sites, in other words the presence
or absence of an effective $\tau$-particle.  Similarly, we argue that
tau desorption, diffusion and absorption can be represented by an
effective undirected motion along the track.\footnote{Of course, this
simplified representation would be inappropriate for addressing
questions about the distribution and dynamics of the tau particles
themselves.  With respect to properties of the kinesin transport, the
validity of the approach will later be further justified by the
similarities between simulation results from this minimal model and
from the full model of Sec.~\ref{s:sim}.}  Hence, in our simplified
model, the $\tau$-particles undergo a symmetric exclusion process
(SEP) in the lower lane.
Specifically, a $\tau$-particle at site $i$ in the lower lane hops
randomly (after an exponentially distributed waiting time) to one of
the nearest-neighbour sites $i-1$ or $i+1$.  The rate for each of
these moves is $D_\tau$ if the destination site is vacant and zero
otherwise.  It is clear that the average occupation number in the
steady state $\rho_\tau \equiv \langle \tau_i \rangle$ is
site-independent.  [Here, as throughout the paper, we use angular
brackets to denote an average over stochastic histories.]

K-particles can occupy sites in both the lower lane and upper lane
(corresponding to kinesin motors being bound or unbound respectively).  Along the
lower lane the dynamics is given by a totally asymmetric exclusion
process with rate $p$ for rightward hops.  Diffusion of motors in the
surroundings of the filament, is represented by a symmetric exclusion
process in the upper lane with rate $D$.  The key feature of the model
(reflecting the underlying biology) is that the coupling of these two
processes depends on the local density of $\tau$-particles.
Specifically, a K-particle in the upper lane hops to a K-vacant
neighbouring site in the lower lane with rate $\omega_a$ if the
destination site is not occupied by a $\tau$ and with a (reduced) rate
$\tilde{\omega}_a$ if there is a $\tau$ there.  In contrast, a
K-particle in the lower lane moves to a vacant neighbouring site in
the upper lane with rate $\omega_d$, regardless of the presence or
absence of $\tau$.

To summarize, our model is defined by the possible moves (and
corresponding rates) listed in Table~\ref{t:moves}.\footnote{Note that
we sometimes use the generic $\omega_x$ for statements which apply
separately to all three inter-lane rates $\omega_a$,
$\tilde{\omega}_a$, $\omega_d$.}
\begin{table*}
\caption{Table of possible moves}
\label{t:moves}
\begin{ruledtabular}
\begin{tabular}{lll}
Move & Rate & Biological interpretation \\
\hline
$\{\tau_i=1, \tau_{i\pm1}=0\} \to \{\tau_i=0,\tau_{i\pm1}=1 \}$ & $D_\tau$ & Effective diffusion of tau along MT \\
$\{b_i=1,b_{i+1}=0 \} \to \{b_i=0,b_{i+1}=1 \}$ & $p$ & Directed motion of kinesin along MT \\
$\{u_i=1,u_{i\pm1}=0 \} \to \{u_i=0,u_{i\pm1}=1 \}$ & $D$ & Diffusion of kinesin in surroundings \\
$\{b_i=1,u_i=0 \} \to \{b_i=0,u_i=1  \}$ & $\omega_d$ & Detachment of kinesin from MT \\
$\{\tau_i=0, b_i=0,u_i=1 \} \to \{\tau_i=0, b_i=1,u_i=0  \}$ & $\omega_a$ & Attachment of kinesin in absence of $\tau$ \\
$\{\tau_i=1, b_i=0,u_i=1 \} \to \{\tau_i=1, b_i=1,u_i=0  \}$ & $\tilde{\omega}_a$ & Attachment of kinesin in presence of $\tau$ \\
\end{tabular}
\end{ruledtabular}
\end{table*}
Even this simple model is difficult to treat analytically since the
K-particles experience a dynamic disorder due to the interaction with
$\tau$.  For reviews of earlier work on disorder in driven diffusive
systems see, e.g.,~\cite{Stinchcombe02,Barma06}.  For a disordered
asymmetric exclusion process with non-conserving sites, Evans~\emph{et
al.}~\cite{Evans04d} were able to solve the steady state exactly in
two limits.  In a similar spirit, we treat below some tractable limits
of the present model as well as discussing more qualitatively the
general case. 

\subsection{Tractable limits}

For a fixed density of $\tau$-particles, the behaviour of the system
is characterized by plotting the stationary current along the lower
lane against the density of K-particles.  In this subsection we show
how this ``fundamental diagram'' can be calculated exactly in certain
limits and compare the $\tau$-induced changes with the observations
from the more realistic model in Sec.~\ref{s:sim}.

\subsubsection{Pure case}
\label{sss:pure}

\begin{figure}
\begin{center} 
\includegraphics*[width=1.0\columnwidth]{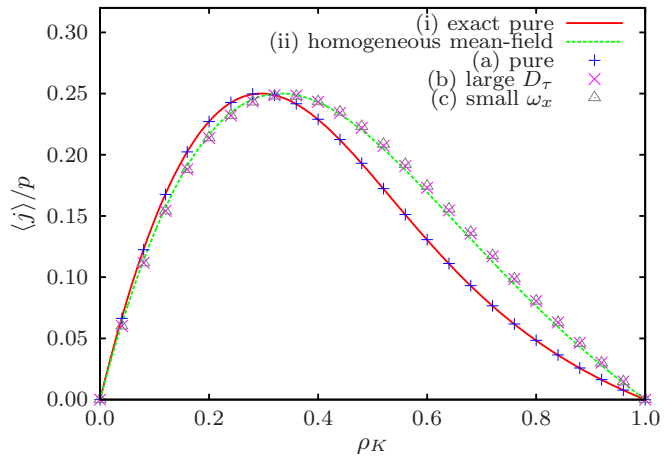}
\caption{(Color online) Current of K-particles (representing kinesin motors) along
  track versus K-density.  Parameters are chosen so that the ratios of
  desorption and absorption rates in the presence and absence of
  $\tau$ are given by $r\equiv \omega_d/\omega_a=0.1$ and $\tilde{r}
  \equiv \omega_d /\tilde{\omega}_a=0.5$.  Lines show (i) the exact
  theoretical result for the pure case ($\rho_\tau=0$) and (ii) the
  simple mean-field result for the case with $\rho_\tau=0.5$.  Points
  denote simulation results from a system with lattice size $L=1000$
  averaged over 1000 realizations and 10000 time steps.
  Individual cases are (a) pure model with parameters: $\rho_\tau=0$,
  $p=0.2$, $D=0.5$, $\omega_d=0.05$, $\omega_a=0.5$; (b) fast $\tau$
  diffusion with parameters: $\rho_\tau =0.5$, $D_\tau=5.0$, $p=0.2$,
  $D=0.5$, $\omega_d=0.05$, $\omega_a=0.5$, $\tilde{\omega}_a=0.01$;
  (c) case of small absorption/desorption rates with parameters
  $\rho_\tau=0.5$, $D_\tau=0.01$, $p=2.0$, $D=5.0$, $\omega_d=0.05$,
  $\omega_a=0.5$, $\tilde{\omega}_a=0.01$. }
\label{f:puremf}
\end{center}
\end{figure}

In the absence of $\tau$ (i.e., $\rho_\tau=0$), the model is exactly
solvable as shown by Klumpp and Lipowsky~\cite{Klumpp03}.  The steady
state is a product state with site-independent K densities in lower
and upper lanes $\rho_b \equiv \langle b_i \rangle$ and $\rho_u \equiv
\langle u_i \rangle$.  Since this stationary state has no correlations
between sites, these densities obey the mean-field equation
\begin{equation}
\omega_d \rho_b(1-\rho_u) = \omega_a \rho_u(1-\rho_b) \label{e:dens}
\end{equation}
which corresponds to the absence of a net current between the two lanes.
Combined with the expression for the (known) total motor density
\begin{equation}
\rho_K = \frac{\rho_b+\rho_u}{2},
\end{equation}
this yields a quadratic equation for $\rho_b$ which can be solved
explicitly.  The mean current $\langle j \rangle$ along the track is
then obtained as
\begin{equation}
{\langle j \rangle} = p \rho_b (1-\rho_b). \label{e:J}
\end{equation}
In Fig.~\ref{f:puremf} we plot $\langle j \rangle/p$ against the
total motor density $\rho_K$; note that in the pure case the shape of
this ``fundamental diagram'' is completely determined by the ratio of
absorption and desorption rates $r\equiv \omega_d/\omega_a$.

\subsubsection{Fast diffusion of $\tau$-particles}
\label{sss:fastt}

If the $\tau$-particles diffuse infinitely fast relative to the
movement of the K-particles then there is no possibility for
correlations to build up in the system and the presence of the
$\tau$-particles merely alters the average absorption rate.  Hence in
the limit $D_\tau \to \infty$ (with the other rates all finite), one
expects that the stationary state is still described by the simple
mean-field theory (i.e., with uncorrelated homogeneous site densities)
with an effective absorption rate given by
\begin{equation}
\omega_{a,\text{eff}}=(1-\rho_\tau) \omega_a + \rho_\tau
\tilde{\omega}_a \label{e:mf}.
\end{equation}
This expectation is confirmed by the simulation results shown in
Fig.~\ref{f:puremf}.  Notice that the effect of the introduction of
$\tau$ (with $\tilde{\omega}_a < \omega_a$) is to shift the position
of the maximum towards higher K-densities.


As an aside, we note that the same mean-field solution is found for
\emph{any} value of $D_\tau$ in the limit of vanishing coupling
between the lanes, $\omega_x \to 0$ (i.e., $\omega_a \to 0$,
$\tilde{\omega}_a \to 0$, $\omega_d \to 0$) with the ratios $r \equiv
\omega_d/\omega_a$ and $\tilde{r} \equiv \omega_d /\tilde{\omega}_a$
constant.  In this limit the time between each lane-changing event
tends to infinity so that after each such event the K-density profiles
in the lower and upper lane relax to the spatially-uncorrelated
homogeneous case and we again recover the mean-field result with
effective rate given by~\eqref{e:mf} (see again the simulation results
in Fig.~\ref{f:puremf}).

\subsubsection{Slow diffusion of $\tau$-particles}
\label{sss:slowt}

\begin{figure}
\begin{center} 
\includegraphics*[width=1.0\columnwidth]{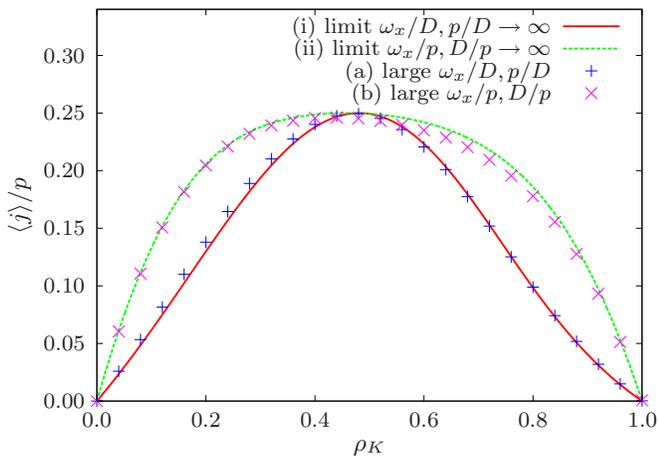}
\caption{(Color online) As Fig.~\ref{f:puremf} but for the quenched disorder limit
  $D_\tau \to 0$.  Lines show theoretical results for (i) infinitely
  fast absorption/desorption and K-movement on track compared to
  K-diffusion in bulk and (ii) infinitely fast absorption/desorption
  and K-diffusion in bulk compared to K-movement on track.  These two
  cases correspond to a homogeneous density profile in the lower and
  upper lanes respectively.  Supporting simulation results are shown
  for parameter values (a) $\rho_\tau=0.5$, $D_\tau=0.01$, $p=2.0$,
  $D=0.01$, $\omega_d=0.5$, $\omega_a=5.0$, $\tilde{\omega}_a=0.1$ and
  (b) $\rho_\tau=0.5$, $D_\tau=0.01$, $p=0.01$, $D=5.0$,
  $\omega_d=0.5$, $\omega_a=5.0$, $\tilde{\omega}_a=0.1$.}
\label{f:quenched}
\end{center}
\end{figure}

Let us now consider the quenched disorder limit, i.e., $D_\tau \to 0$.
In this case, the time between movement of $\tau$-particles tends to
infinity and between these events the system relaxes to a
quasi-stationary state with site-dependent densities of K-particles.
There is a growing body of literature discussing lattice gas models
with quenched spatial disorder but analytical progress is difficult
even for the usual single-lane ASEP (see, e.g.,~\cite{Tripathy98,
Krug00,Kolwankar00, Me04, Juhasz05b}).  For our two-lane model, we
focus here on two limiting cases where the analysis again becomes relatively
straightforward;
comparison with simulation is shown in Fig.~\ref{f:quenched}.


Firstly, we consider taking $D_\tau \to 0$ followed by
$\omega_x/D,p/D \to \infty$ whilst holding the other ratios between rates
constant.  Here the density profile in the lower lane relaxes rapidly
to stationarity between each hopping event in the upper lane.  In
other words, the profile in the lower lane is that of a homogeneous
ASEP (constant density $\rho_b$) whereas the density in the upper lane
is inhomogeneous and determined by the local absorption and desorption
rates.  Specifically, there are two distinct unbound densities
$\rho_u$ and $\tilde{\rho}_u$ corresponding to the absence and
presence of $\tau$ respectively.  These densities must obey the
relations
\begin{eqnarray}
r \rho_b(1-\rho_u) &= \rho_u(1-\rho_b) \label{e:densa} \\
\tilde{r} \rho_b(1-\tilde{\rho}_u) &= \tilde{\rho}_u(1-\rho_b). \label{e:densb}
\end{eqnarray}
and the equation for total K-particle density becomes
\begin{equation}
\rho_K = \frac{1}{2}\left[\rho_b + (1-\rho_\tau)\rho_u  + \rho_\tau \tilde{\rho}_u\right]. \label{e:norm}
\end{equation}
Substituting~\eqref{e:densa} and~\eqref{e:densb} into~\eqref{e:norm} gives
\begin{equation}
\rho_K =  \frac{1}{2}\left[\rho_b + \frac{r (1-\rho_\tau) \rho_b}{1+(r-1)\rho_b} + \frac{\tilde{r} \rho_\tau \rho_b}{1+(\tilde{r}-1)\rho_b}\right]. 
\end{equation}
For fixed $\rho_K$ this is a cubic equation for $\rho_b$; numerical
solution followed by substitution into~\eqref{e:J} yields a current in
good agreement with simulation.\footnote{Even better agreement would
  presumably be obtained for $D_\tau$ smaller relative to $D$ but then
  longer simulation times would be required to reach the steady
  state.}


Secondly we consider the case where $D_\tau \to 0$ and then
$\omega_x/p, D/p \to \infty$ (again with the other ratios between
rates held constant).  Here the density in the upper lane relaxes to
that of a homogeneous SEP (constant density $\rho_u$) and the density
in the lower lane is inhomogeneous with two possible values $\rho_b$
and $\tilde{\rho}_b$ given by the local absorption and desorption
rates via
\begin{eqnarray}
r \rho_b(1-\rho_u) &= \rho_u(1-\rho_b) \label{e:dens1} \\
\tilde{r} \tilde{\rho}_b(1-\rho_u) &= \rho_u(1-\tilde{\rho}_b). \label{e:dens2}
\end{eqnarray}
These densities must satisfy
\begin{equation}
\rho_K = \frac{1}{2}\left[(1-\rho_\tau)\rho_b  + \rho_\tau \tilde{\rho}_b + \rho_u\right]
\end{equation}
leading to
\begin{equation}
\rho_K = \frac{1}{2}\left[\frac{(1-\rho_b^\tau) \rho_u}{r+(1-r)\rho_b} + \frac{\rho_b^\tau \rho_u}{\tilde{r}+(1-\tilde{r})\rho_u} + \rho_u\right].
\end{equation}
After solving this cubic equation for $\rho_u$ one can obtain the two
bound densities via~\eqref{e:dens1} and~\eqref{e:dens2}.  The current
along the lower lane is of order $p$ and its average value can be calculated
by using the probabilities
to find a $\tau$-particle at either end of a given bond:
\begin{multline}
\langle j \rangle = p[(1-\rho_\tau)^2 \rho_b(1-\rho_b) + \rho_\tau
  (1-\rho_\tau) \tilde{\rho}_b(1-\rho_b) \\ + (1-\rho_\tau) \rho_\tau \rho_b(1-\tilde{\rho}_b) + (\rho_\tau)^2 \tilde{\rho}_b(1-\tilde{\rho}_b)]. \label{e:Jin}
\end{multline}
Of course, on very long timescales the positions of the $\tau$-particles and
hence the local densities of K-particles will change.  However, the
system will rapidly relax to a new quasi-stationary state which, on
average, has the same K-current (as follows from conservation of
$\tau$-particle number).  The prediction of~\eqref{e:Jin} is again
well-supported by simulation even for parameter values, which for
computational efficiency, are still relatively far from the limiting
case; see Fig.~\ref{f:quenched}.

\subsection{Intermediate parameters}

\begin{figure}
\begin{center} 
\includegraphics*[width=1.0\columnwidth]{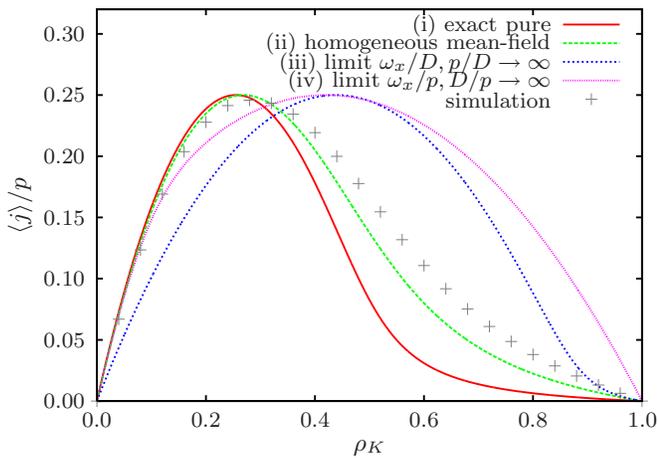}
\caption{(Color online) Current of K-particles along track versus total K-density per
  unit length for $\rho_\tau=0.75$, $D_\tau=0.01$, $p=0.5$, $D=1.0$,
  $\omega_d=0.01$, $\omega_a=1.0$, $\tilde{\omega}_a=0.01$.  Crosses
  show simulation results (parameters as in
  Fig.~\ref{f:puremf}); lines show analytical results for the
  limiting cases of~\ref{sss:pure}--\ref{sss:slowt}.} 
\label{f:eg2}
\end{center}
\end{figure}

All the limits treated in the previous subsection correspond to cases
where in the (quasi-)stationary state there are no correlations
between the occupations of different sites (i.e., they are situations
which can be described by forms of mean-field theory).  However, we
emphasize that this is not expected to be the case for general choices
of rates.  In this subsection we present illustrative simulation
results for intermediate parameters---see Fig.~\ref{f:eg2}.

The relative values chosen are inspired by the biological context
 although we make no attempt to quantitatively match our simple model
 to real data.  Specifically, the bulk diffusion and absorption rate
 are large compared to the hopping rate on the track but the
 desorption rate is small.  The presence of tau strongly reduces the
 absorption probability ($\tilde{\omega}_a \ll
 \omega_a$)~\cite{Seitz02} while the effective diffusion rate for tau
 is rather small.  The simulation data shown in Fig.~\ref{f:eg2} shows
 that, just as in the limiting cases discussed
 in~\ref{sss:fastt}--\ref{sss:slowt}, there is a shift in the position
 of the maximum towards higher K-densities (compared to the pure
 case).  However, significantly, there is also a decrease in the
 height of this maximum.  We remark that a shift in the maximum, a
 reduction in its height and significant correlations were all
 observed in simulations of the more realistic model in
 Sec.~\ref{s:sim}.

From a theoretical point of view, the shape of the fundamental diagram
is interesting (in particular, the effect of tau on the position and
height of the maximum).  Physiologically however, one expects the
number of kinesin motors to be approximately constant, and thus a more
relevant question is the dependence of current on tau concentration at
a fixed low density of motors. Of course, it is possible to
analytically obtain this dependence in the tractable limits discussed
above.  For more ``realistic'' intermediate parameters, a good
approximate description in the low K-density regime can be obtained by
considering the random walk properties of single motors which is the
approach we choose to pursue in the following section.  To simplify
the analysis while retaining the most important features we henceforth
set $D_\tau=0$ and $\tilde{\omega}_a=0$.\footnote{We can appeal to
physical intuition and simulation results
(cf.~Fig.~\ref{f:quenched}--\ref{f:eg2}) to support the assertion
that, with respect to the current of $K$-particles, there is no
singular behaviour at this limit.}

\section{Tau-dependence of current}
\label{s:tauvary}

\subsection{Single-motor properties}

In this subsection we show how the properties of a single motor (or
equivalently an ensemble of non-interacting motors) can be
analytically understood through a random walk picture.  This
facilitates an investigation of the effect of tau on the average
velocity and diffusion constant of the motor (or equivalently the mean
and variance of the current).  Such a single-particle random walk
approach (cf.~\cite{Ajdari95,Lipowsky01,Nieuwenhuizen02}) is relevant for
\emph{in vitro} experiments involving individual motors but, as we
shall see in~\ref{ss:lowmd}, also gives valuable
information about the low-motor-density limit of the interacting-motor
problem.

\subsubsection{Probability distributions, large deviations}
Let us start by considering a single motor moving in the quenched
disordered landscape defined by Fig.~\ref{f:model} with $D_\tau=0$
and $\tilde{\omega}_a=0$.  We are interested in the distribution of
the total number of steps $Y_1(t)$ made by a single motor in the lower
lane over some time period $[0,t]$ (i.e., the integrated current in
time $t$).  Now it is obvious that, if the motor spends a fraction $x$
of its time in the lower lane, then $Y_1$ has a Poisson distribution
with mean $pxt$, i.e.,
\begin{equation}
P_x(Y_1,t)=\frac{ e^{-pxt} (pxt)^{Y_1}}{Y_1 !}
\end{equation}
or equivalently the observed time-averaged current $y_1 \equiv Y_1/t$
has the distribution
\begin{equation}
P_x(y_1,t)=\frac{ e^{-pxt} (pxt)^{y_1 t}}{(y_1 t)!}.
\end{equation}
The full distribution of $y_1$ is then obtained by averaging over all possible values of $x$, i.e.,
\begin{equation}
P(y_1,t)=\int_0^1 P_x(y_1,t) P(x,t) \, dx, \label{e:jint}
\end{equation}
where $P(x,t)$ is the distribution of the fraction of time (from a total $t$) spent in the lower lane.
Our central task is thus to calculate this distribution and hence obtain the distribution of $y_1$.  We
note that $P(y_1,t)$ is expected to have a large deviation form
\begin{equation}
P(y_1,t) \sim e^{-t\hat{e}(y_1)} \label{e:pj} 
\end{equation}
and that knowledge of the large deviation function $\hat{e}(y_1)$
gives the asymptotic behaviour of all moments of the distribution.

Before showing how to obtain $P(x,t)$ approximately in the disordered
case, we first digress to discuss the pure result corresponding to
$\rho_\tau=0$.  In this case, the upper and lower lanes form an
effective two-state Markovian system with rates $\omega_a$ and
$\omega_d$ for transitions between the states.  It can then be shown
exactly that the distribution of the total fractional time in the
lower (bound) state is given by~\cite{Berezhkovskii99}
\begin{multline}
P(x,t)=\delta(1-x)e^{-(\omega_a + \omega_d) P^0_u t} +
(\omega_a+\omega_d) P^0_u  t \times \\\left\{ I_0(X) + \sqrt{\frac{x P^0_b}{(1-x) P^0_u}} I_1(X) \right\} e^{-[ x P^0_u + (1-x) P^0_b] (\omega_a+\omega_d) t}\label{e:ppx}
\end{multline}
where $\delta$ is the Dirac delta function, $I_0$ and $I_1$ are Bessel functions, $P^0_b$ and $P^0_u$ are the probabilities to find the motor in the bound and unbound states respectively and
\begin{equation}
X \equiv 2(\omega_a + \omega_d) t \sqrt{P^0_b P^0_u x(1-x)}.
\end{equation}
Furthermore, it is trivial to show that
\begin{equation}
P^0_b = \frac{\omega_a}{\omega_a + \omega_d}, \qquad P^0_u = \frac{\omega_d}{\omega_a + \omega_d}.
\end{equation}
The expression~\eqref{e:ppx} can then be substituted
into~\eqref{e:jint} and numerical evaluation of the resulting integral
agrees with the current distribution obtained by
simulation (not shown).
In the long-time limit~\eqref{e:ppx} reduces to a Gaussian~\cite{Geva98}. 

\subsubsection{Disordered model as trapping problem}
In the disordered case (non-zero density of tau) the upper and lower
lanes no longer form an effective two-state Markovian system (the
average rate for escape from the upper lane is dependent on the
occupation time, since for longer times the particle is more likely to
be found trapped in long tau-decorated regions).  However, for a
two-state non-Markovian system, it can be shown~\cite{Boguna00} that
the asymptotic distribution of occupation times is still Gaussian with
the fraction $x$ of time in the lower lane having mean
\begin{equation}
P_b=\frac{\langle t_b \rangle}{\langle t_b \rangle + \langle t_u \rangle}, \label{e:Pbp}
\end{equation}
and variance given by $\Delta/t$ where
\begin{equation}
{\Delta} = \frac{\sigma_u^2 \langle t_b \rangle^2 + \sigma_b^2 \langle t_u \rangle^2}{(\langle t_b \rangle + \langle t_u \rangle)^3}. \label{e:Deltap}
\end{equation}
Here $\langle t_u \rangle$ ($\langle t_b \rangle$) and $\sigma_u^2$
($\sigma_b^2$) are the mean and variance of the sojourn time
distribution in the upper (lower) lane.  Since the rate for exiting
the lower lane is always $\omega_d$ we have
\begin{equation}
\langle t_b \rangle = \frac{1}{\omega_d}, \qquad
\sigma_b^2=\frac{1}{\omega_d^2} \label{e:tb}
\end{equation}
and the only remaining difficulty is to calculate the equivalent
quantities for a sojourn in the upper lane.  Obviously, $\langle t_u
\rangle$, $\langle t_u^2 \rangle$ and hence $P_b$ and $\Delta$ are
functions of $\rho_\tau$ but for simplicity we suppress this dependence
notationally.

Now, focusing in on the upper lane, we see that a K-particle performs an
ordinary random walk (with hopping rate $D$) in regions of sites
occupied by tau.  When it reaches a site without tau, it has a
probability $A = \omega_a/(2D+\omega_a)$ to be ``trapped'' and move to
the lower lane.  For several decades such one-dimensional trapping
problems have been extensively analysed in the literature (see,
e.g.,~\cite{Weiss85,VandenBroeck86,Haus87} and references therein).
In particular it can be shown (see,
e.g.,~\cite{Nieuwenhuizen90}) that the tail of the sojourn time
distribution has a stretched exponential form $P(t_u) \sim \exp(- a
\left(\ln (1/\rho_\tau) \right)^{2/3} t_u^{1/3}) $  due to the survival of particles in
arbitrarily large trap-free regions.  However, we are interested in
the mean and variance which are not much influenced by the tail
of the distribution.  Here we show how, within certain
approximations, these quantities can be calculated directly.  Our
approach is in the spirit of a continuous time-version of the method
given in~\cite{Hatlee81}.

\begin{figure}
\begin{center} 
\includegraphics*[width=0.9\columnwidth]{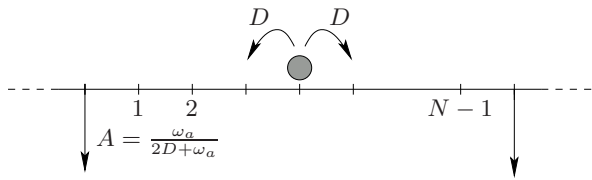}
\caption{Schematic of a K-particle diffusing on the $N-1$
  tau-occupied sites between two tau-free sites (traps).  This can be considered as a trapping reaction with trapping probability $A= \omega_a / (2D+\omega_a)$.}
\label{f:trap}
\end{center}
\end{figure}

First we consider two traps separated by a ``line segment'' of length
$N$ as shown in Fig.~\ref{f:trap}.  
In other words, we have a tau-free site at $l=0$ followed by $N-1$
sites occupied by tau, followed by another tau-free site at $l=N$.
The time to reach one of the traps at either end starting from
position $l$ is a random variable $T_l(N)$; it can be 
shown by standard arguments~\cite{Weiss81} that its mean is given by
\begin{equation}
\langle T_l(N) \rangle =\frac{(N-l)l}{2D}. \label{e:TlN}
\end{equation}
Note that this expression is invariant under the transformation $l
\leftrightarrow N-l$ as expected from the symmetric nature of the
diffusion.  We now make the important assumption that the distribution
of entry points into the upper lane has a uniform distribution.  This
is expected to be a good approximation when the typical run length in
the lower lane is large compared to the spacing of traps, i.e., when
\begin{equation}
\frac{p}{\omega_d} \gg \frac{1}{1-\rho_\tau}.
\end{equation}
Using $T(N)$, without subscript, to denote the time to absorption from a random initial position we then obtain
\begin{align}
\langle T(N) \rangle &= \frac{1}{N} \sum_{l=1}^{N-1} \langle
T_l(N) \rangle \\
&= \frac{N^2 - 1}{12 D} \label{e:TN}
\end{align}
Following~\cite{Weiss81}, one can also write down expressions (see appendix) for the
mean square first passage times $\langle T_l(N)^2 \rangle$ and
$\langle T(N)^2 \rangle \equiv (1/N) \sum_{l=1}^{N-1} \langle T_l(N)^2
\rangle$.  Combined with a known distribution of traps
(i.e., the distribution of lengths $N$), this is sufficient to obtain
explicit expressions for the first two sojourn time moments in the
case where the trapping probability $A$ is unity (i.e., $\omega_a \to
\infty$).  However, for imperfect trapping one must allow for the
possibility that the particle is not absorbed on its first visit to
the trap.  We now discuss two possible approximations which allow
further analytical progress:

\begin{enumerate}

\item In the ``reflecting trap'' or ``decoupled-ring
  approximation''~\cite{Weiss85, Aspelmeier02} it is assumed that a
  particle which is not absorbed by a trap is \emph{always} reflected
  back into the same trap-free line segment from which it
  came.\footnote{A particle which enters the upper lane at a trap site
  is randomly assigned to the line segment to the left or the
  right.}
  This simplifies the analysis by decoupling the tau segments between
  traps but at the expense of introducing an uncontrolled
  approximation.  The approach is clearly exact in the case of perfect
  traps (i.e., in our model for $\omega_a \to \infty$) but has also
  been shown to give good results for trapping probabilities smaller
  than unity~\cite{Aspelmeier02}. \label{num:reflect}

\item Another approach is to place each particle which escapes from a
  trap site on to the first site (i.e., $l=1$) in a line segment
  with length $N$ randomly drawn from the distribution of all possible
  lengths.  Again this approximation is exact for $\omega_a \to
  \infty$.  Furthermore, we might also expect it to be exact in the limit of
  zero-trapping probability ($\omega_a \to 0$) since in this case each
  particle explores the full distribution of tau line
  segments. \label{num:un}

\end{enumerate}
Whereas the first approximation treats the line segments as
unconnected, the second assumes they are \emph{all} interconnected,
in the sense that a particle can escape from a trap into \emph{any}
line-segment (rather than just neighbouring segments as happens in
reality).  For intermediate trapping probabilities, we might expect
the true distribution of first passage times to fall somewhere in
between the predictions of these two limiting cases.
The analytical approach proceeds in a similar way for both
approximations; we give here the details for approximation~\ref{num:reflect},
indicating only briefly the modifications necessary for approximation~\ref{num:un}.

Let us start by considering a particle which enters the upper lane at
a random position in a tau segment of length $N$.  The total sojourn time
spent by the particle in the upper lane $t_u(N)$ is a random variable
given by the following sum
\begin{equation}
t_u(N) = T(N) + \sum_{i=1}^{m+1} T_\text{tr}^{(i)} + \sum_{i=1}^m
T_1^{(i)}(N_i). \label{e:tuN}
\end{equation}
Here $m$ is the number of times the particle is reflected from the
trap, $T(N)$ is the time to reach the trap for the first time
(from a random initial position), $T_\text{tr}^{(i)}$ is the time
spent at the trapping site on the $i$th visit and $T_1^{(i)}(N_i)$ is
the time to return to the trap after escaping for the $i$th time.
Under approximation 1, the particle is always reflected back into the
same line segment and $N_i=N$.  Averaging over stochastic histories
then yields
\begin{equation}
\langle t_u(N) \rangle_1  = \langle T(N) \rangle + \langle m  +
1 \rangle \langle T_\text{tr} \rangle + \langle m \rangle \langle T_1(N)
\rangle
\end{equation}
where we have used the independence of the various random variables in
order to factorize the expectation values.

Noting that the number of reflections has a geometric distribution
given by
\begin{equation}
P(m)=A(1-A)^m 
\end{equation}
we immediately find
\begin{equation}
\langle m \rangle = \frac{1-A}{A}. 
\end{equation}
Furthermore the time spent at the trap is exponentially distributed
with mean
\begin{equation}
\langle T_\text{tr}\rangle = \frac{1}{2D+\omega_a}. 
\end{equation}
Together with~\eqref{e:TlN} and~\eqref{e:TN}, this allows a complete
determination of $\langle t_u(N) \rangle_1$.

Next we note that the probability of entering the upper lane in a
given line segment (i.e., a sequence of consecutive sites occupied by
tau) is proportional to its length.  Hence for a uniform distribution
of $\tau$-particles we should average $\langle t_u(N) \rangle_1$ over
the distribution
\begin{equation}
P_0(N)= N \rho_\tau^{N-1} (1-\rho_\tau)^2.
\end{equation}
The resulting infinite sum
\begin{equation}
\langle t_u \rangle_1 = \sum_{N=1}^\infty P_0(N) \langle t_u(N) \rangle_1,
\end{equation}
contains summands proportional to $N^\alpha \rho_\tau^N$ with $\alpha=1,2,3$
and evaluates to
\begin{equation}
\langle t_u \rangle_1 =\frac{\rho_\tau}{2D (1-\rho_\tau)^2} +
\frac{1+\rho_\tau}{\omega_a (1-\rho_\tau)}. \label{e:tu}
\end{equation}
The only difference in approximation~\ref{num:un} is that the lengths
$N_i$ appearing in~\eqref{e:tuN} are drawn from the distribution
$P(N_i)=\rho_\tau^{N_i-1} (1-\rho_\tau)$.  After averaging over this
distribution as well as the initial distribution one finds
\begin{equation}
\langle t_u \rangle_2 =\frac{\rho_\tau}{2D (1-\rho_\tau)^2} +
\frac{1}{\omega_a (1-\rho_\tau)}.
\end{equation}

In similar fashion one can obtain an expression for $\langle
t_u(N)^2 \rangle$ from~\eqref{e:tuN}.  The resulting calculations are
straightforward but tedious and are outlined for convenience in the
appendix.  Using approximation~\ref{num:reflect} one eventually
obtains
\begin{multline}
\langle t_u^2 \rangle_1 =  \{ \omega_a^2
  \rho_\tau(1+\rho_\tau)^2  +
  8D\omega_a\rho_\tau(1-\rho_\tau^2)
\\ +4D^2(1-\rho_\tau)^2(1+4\rho_\tau+\rho_\tau^2)
  \}
\\
  \times \{2(1-\rho_\tau)^4 D^2 \omega_a^2\}^{-1}. \label{e:tu2}
\end{multline}
Results~\eqref{e:tu} and~\eqref{e:tu2} together
with~\eqref{e:Pbp}--\eqref{e:tb} finally determine the parameters of the
asymptotic Gaussian form of $P(x,t)$.

\subsubsection{Results for velocity and diffusion coefficient}

After the effort involved in finding the long-time behaviour of
$P(x,t)$, it is comparatively straightforward to evaluate the
integral~\eqref{e:jint}---one simply uses Stirling's approximation for
the factorials in the Poisson distribution $P_x(y_1,t)$ and then
employs a saddle-point treatment.  This finally yields the large
deviation function of the observed average current along the lower
lane
\begin{equation}
\hat{e}(y_1) = -\frac{y_1}{2} + y_1\ln\frac{y_1}{p x_+} + \frac{(p\Delta-P_b)x_+}{2\Delta} + \frac{P_b^2}{2\Delta} \label{e:larged}
\end{equation}
where $x_+$ is the positive solution of the quadratic equation
\begin{equation}
x^2 + (p\Delta-P_b)x-y_1\Delta=0.
\end{equation}
The large deviation function gives knowledge of all cumulants of
$Y_1$.  In particular, we easily obtain the mean velocity
\begin{align}
v_1 &\equiv \lim_{t\to\infty} \frac{\langle Y_1 \rangle}{t} \\
&=p P_b
\end{align}
and diffusion constant
\begin{align}
D_1 &\equiv \lim_{t \to \infty} \frac {\langle Y_1^2 \rangle - \langle Y_1 \rangle^2 }{t} \\
&= pP_b + p^2 \Delta.
\end{align}
In fact, this is simply the expected result for the variance of a
compound process.  However, we emphasize that
the large deviation expression~\eqref{e:larged} also gives complete
information about the asymptotic behaviour of higher moments.

\begin{figure}
\begin{center}
\includegraphics*[width=1.0\columnwidth]{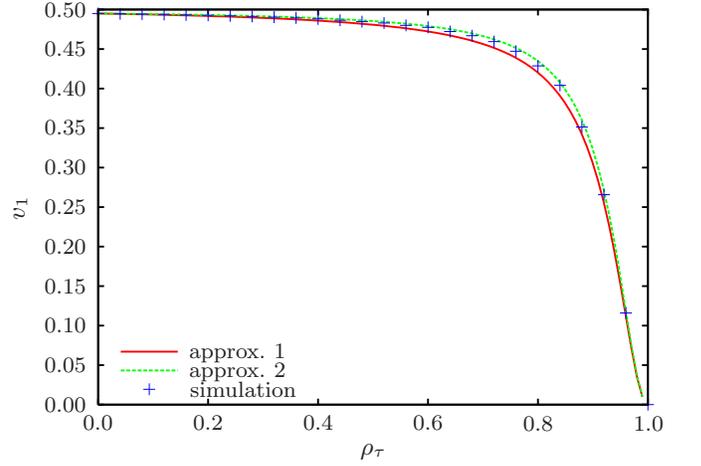}
\caption{(Color online) Comparison of analytical and simulation results for the
  dependence of single motor velocity (along microtubule) on
  concentration of tau.  Parameter values are $p=0.5$, $D=1.0$,
  $\omega_a=1.0$ and $\omega_d=0.01$.  Simulation results are for a
  system of size $L=10000$ with a fixed realization of disorder for
  each density value; averages are taken over 100000 time steps and
  10000 histories. 
}
\label{f:singmean}
\end{center}
\end{figure}

Using~\eqref{e:Pbp} and~\eqref{e:Deltap} for the bound-state
occupation probability and its variance, we obtain explicit
expressions for velocity and diffusion.  Specifically, the observed
average motor velocity along the lower lane in the presence of a
quenched distribution of $\tau$ is given by (under approximation 1)
\begin{equation}
\langle v_1 \rangle_1 =\frac{p(1-\rho_\tau)^2}{(1-\rho_\tau)^2 + \left(\frac{\omega_d}{2D}\right)\rho_\tau + \left(\frac{\omega_d}{\omega_a}\right) (1-\rho_\tau^2)}. \label{e:jsinfin}
\end{equation}
In Fig.~\ref{f:singmean} we compare the analytical predictions of
both approximate schemes with simulation
results for the mean velocity of a single motor. 
We find that both approximations reproduce well the decrease of the
average velocity with increasing tau concentration and that, as anticipated, the simulation results lie between the two curves.

\begin{figure}
\begin{center}
\includegraphics*[width=1.0\columnwidth]{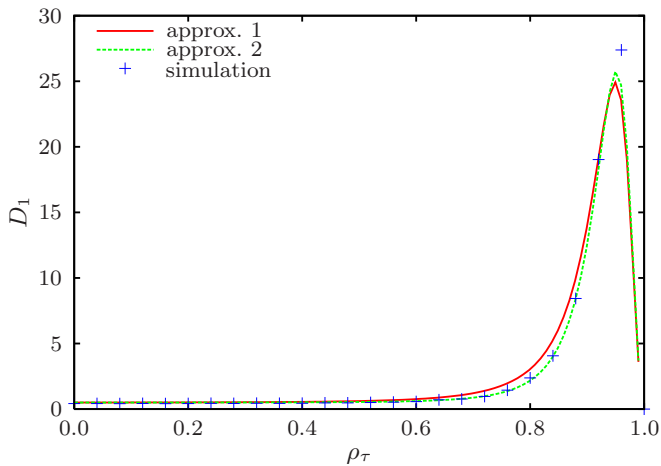}
\caption{(Color online) Comparison of analytical and simulation results for the
  dependence of single motor diffusion constant (along microtubule) on
  concentration of tau.  Parameters as in Fig.~\ref{f:singmean}.
}
\label{f:jsinvar}
\end{center}
\end{figure}

Similarly the analytically-calculated diffusion constant $D_1$
compares well with simulation, as shown in Fig.~\ref{f:jsinvar}.
In particular, we see that although the system is relatively robust to
low concentrations of tau, for high concentrations there is a dramatic
increase in the variance which could be physiologically significant.

\begin{figure}
\begin{center}
\includegraphics*[width=1.0\columnwidth]{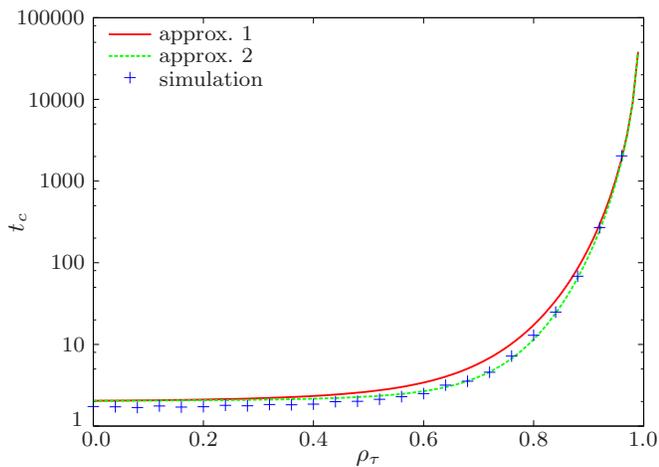}
\caption{(Color online) Crossover timescale $t_c \equiv D_1/(v_1)^2$ for parameter
  values $p=0.5$, $D=1.0$, $\omega_a=1.0$ and $\omega_d=0.01$. Parameters as in Fig.~\ref{f:singmean}.
 Note
  logarithmic scale. 
}
\label{f:crossover}
\end{center}
\end{figure}

In fact, it is perhaps more illuminating to consider the diffusion
constant divided by the square of the velocity as shown in
Fig.~\ref{f:crossover}. 
This defines a crossover timescale $t_c \equiv D_1/(v_1)^2$.  For
times less than $t_c$, diffusive fluctuations dominate whereas for
larger times the directed motion is dominant. 

\subsection{Low motor density}
\label{ss:lowmd}

The single motor calculations of the previous subsection can be used
to give approximate results for the low-motor-density limit of the
full model.  The key observation is that, for realistic parameters,
the total unbound density is much smaller than the total bound density
(except for $\rho_\tau$ very close to unity).  It is thus
reasonable to neglect the exclusion interaction both in the upper lane
and also for inter-lane moves.  In other words we have an asymmetric
exclusion process on the filament coupled to random walks (without
exclusion) in the surroundings. 

\begin{figure}
\begin{center}
\includegraphics*[width=1.0\columnwidth]{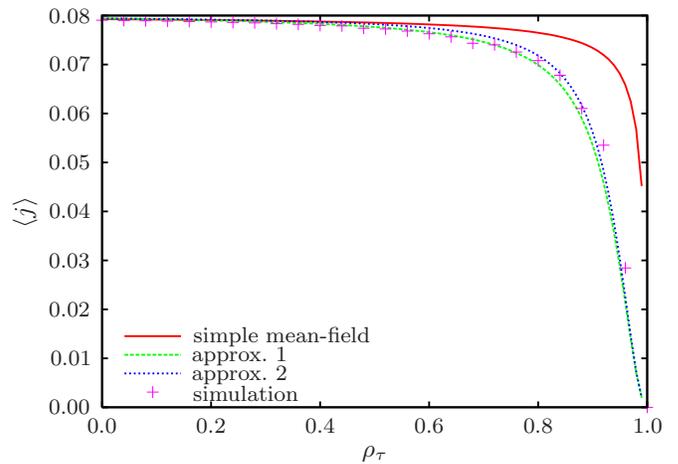}
\caption{(Color online) Comparison of analytical and simulation results for
  mean current of K-particles along the lower-lane (microtubule)
  versus concentration of tau.  Motor density is fixed at $\rho_K=0.1$
  and hopping rates are $p=0.5$, $D=1.0$, $\omega_a=1.0$ and
  $\omega_d=0.01$.  Simulation results are for a system of size
  $L=1000$ with a fixed realization of disorder for each density
  value; averages are taken over 10000 time steps and 100
  histories.}
\label{f:multmean}
\end{center}
\end{figure}

\begin{figure}
\begin{center}
\includegraphics*[width=1.0\columnwidth]{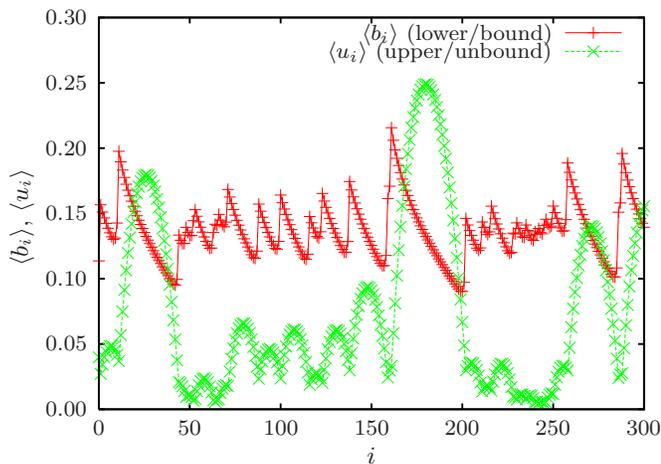}
\caption{(Color online) Site-dependent motor densities $\langle
  b_i \rangle$ and $\langle u_i \rangle$ in lower and upper lanes of the system studied in
  Fig.~\ref{f:multmean} with $\rho_\tau=0.9$.  The data is again averaged over
  1000 histories with a fixed realization of disorder; lines are
  provided as an aid to the eye.}
\label{f:profs}
\end{center}
\end{figure}

In this approximation one then assumes a homogeneous bound density given by 
\begin{equation}
\rho_b= 2\rho_K \times P_b.
\end{equation}
In other words a proportion $P_b$ of the total K-particles are expected
to be in the lower lane (where $P_b$ is the bound-state occupation
probability calculated for the single motor case).  Assuming no
correlations between the bound densities on neighbouring sites, the associated
mean particle current is given by
\begin{equation}
\langle j \rangle = 2 p \rho_K P_b (1 - 2 \rho_K P_b).
\end{equation}
As seen in Fig.~\ref{f:multmean}, 
this approach is a considerable improvement over the simple mean-field
(i.e., neglecting all correlations and simply replacing $\omega_a$ by the effective absorption
rate given by~\eqref{e:mf})
since it implicitly takes account of correlations between tau and
unbound motor density.  However, it still assumes an uncorrelated
uniform density of bound motors (only expected to be a good
approximation when many hops along the filament take place between
absorption/desorption events).  To test this assumption we plot, in
Fig.~\ref{f:profs}, 
averaged density profiles for the rate parameters used in
Fig.~\ref{f:multmean} with $\rho_\tau=0.9$.  In fact there \emph{is}
significant spatial dependence in the bound density but the relative
variation is much less than in the unbound density (where we find
peaks corresponding to the trapping of particles in tau-occupied
regions).  As a quantitative comparison we remark that for the
scenario shown in Fig.~\ref{f:profs} the standard deviation of the
lower-lane densities is approximately 0.14 of their mean (over 1000
lattice sites) whereas the equivalent ratio for the upper lane is
about 0.84.  For larger $p$ the run length of the molecular motors
increases relative to the trap spacing so that the bound density
becomes much more homogeneous while the unbound density is essentially
unchanged (not shown).

\begin{figure}
\begin{center}
\includegraphics*[width=1.0\columnwidth]{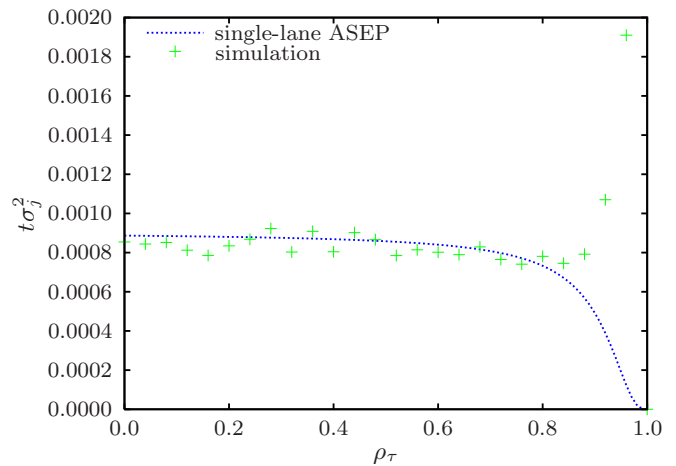}
\caption{(Colour online) Simulation results for variance $\sigma_j^2=\langle j^2
  \rangle - \langle j \rangle^2$ in current of K-particles along the
  lower-lane (microtubule) versus concentration of tau.  Parameters as
  in Fig.~\ref{f:multmean}.
  Dotted line shows expected variance for single-lane ASEP
  with density $2 \rho_K P_b$; peak at high tau-density is
  attributed to inter-lane fluctuations.}
\label{f:multvar}
\end{center}
\end{figure}

Based on the single motor results we also expect high concentrations
of tau to significantly increase the variance of the bound K-particle
current.  Indeed, this is observed in simulations as shown in
Fig.~\ref{f:multvar}.
For low tau-densities we see a plateau where the variance is dominated
by that of the ordinary single-lane exclusion process, which can be calculated
exactly~\cite{Derrida93,Derrida98}.  At a high tau concentration, a
peak appears resulting from the inter-lane fluctuations (just as in
the single motor case).  It seems difficult to obtain the shape of this
peak analytically since fluctuations along the lower lane and
fluctuations between the lanes are \emph{not} independent.

\section{Discussion}
\label{s:diss}

This work is concerned with the stochastic modelling of
molecular-motor traffic inside cells.  In particular, we were motivated by experimental observations of the influence of the
linker protein tau on transport in axons~\cite{Trinczek99,Seitz02}.
These experiments suggest that the primary effect of 
tau proteins is to reduce significantly the binding-affinity of the
motor proteins rather than to alter their dynamics on the
microtubules.  

In order to model this situation, we extended an earlier lattice-based
approach due to Klumpp and Lipowsky~\cite{Klumpp03}.  Specifically,
whereas the original model contained only a single species of particles
(representing kinesin motors), we introduced a second type of particle
to represent the tau proteins.  Following Klumpp and Lipowsky we
modelled the axon in a simplified fashion using a one-dimensional
track embedded in a cylindrical geometry.  The dynamics of the model was defined
to reflect the picture suggested by experiment so that the local presence of
tau reduces the rate for kinesin particles to attach at a potential binding
site but does not affect the rates for movement along the track.

Simulation results for this model show that for a wide range of
parameters, the presence of tau causes only a marginal reduction in
the \emph{maximum} transport along the axonal filament.  (This finding
is consistent with the very recent results of~\cite{Chai09}.)  However
the position of the maximum is shifted to higher kinesin densities and
for a given density of motor proteins one may observe a vigorous
alteration of the flow. This result indicates that an overexpression
of tau proteins may perturb the intracellular transport significantly.
The central aim of the present work was to characterize and
understand this disruption within simple models and thus gain
tentative insight into possible mechanisms in the real biological
context.  (We have not attempted to make quantitative predictions and
further work would be needed to determine to what extent our
observations are seen in real physiological conditions.)

In this spirit, in order to analyse in more detail the origin of the
transport disruption we further simplified the modelling approach by
introducing a two-lane model, which is analytically treatable for
selected, but generical, limiting cases.  For the case of the pure
model, i.e., the case without ``$\tau$-particles'', it is known that
the stationary distribution is given by a product
measure~\cite{Klumpp03}.  However, with the introduction of
$\tau$-particles, mean-field properties are only found in certain
limits.  For example, the case of fast diffusion of $\tau$-particles
can be described by introducing effective binding rates for the
kinesin-particles.

A qualitatively different behaviour is observed if we consider slow
diffusion of $\tau$-particles. This corresponds to quenched disorder on the
track leading to non-trivial density profiles of the ``K-particles'' which represent the motors. For arbitrary
model parameters this limit is difficult to analyse. However, there
are two limiting choices of motor rates which still yield a product
measure for the motor distribution---these choices correspond to a
space-dependent distribution of particles in one lane and a
homogeneous distribution in the other.  In these cases one observes
that the position of the maximal current can be significantly changed by the $\tau$-particles.
A shift of the maximum to higher K-densities (for fixed
$\tau$-densities) is also a feature of simulation results for more
general parameters.  However, when correlations cannot be neglected
one also sees a decrease in the height of the maximum.  In these
respects the simplified two-lane model reproduces to a large extent
the properties of the full model.

In the context of the biological system, it is instructive to analyse the
features of the model for a fixed density of motor proteins and different
concentrations of tau-proteins.   Within the two-lane setup we were
able to obtain analytical results for the random walk
behaviour of a \emph{single} K-particle moving among a fixed background
of $\tau$-particles.  This approach also gives approximate information
about the low-motor-density limit of the simplified model.  The calculations and supporting simulations demonstrate that
for a wide range of
$\tau$-densities (with fixed motor density), the current on the filament remains unchanged. At high
$\tau$-concentrations, however, one observes a drastic decrease of the
system's capacity in a rather small interval of $\tau$-densities. In addition,
the fluctuations of the current show a pronounced maximum, such that the
crossover-times, which characterize the transition from diffusive to
directed motion diverge. 

It should be emphasized that these results were obtained in the
context of a simplified ``toy'' model and caution must be applied in
drawing conclusions about real physiological systems.  This is
particularly true because we assumed throughout a strong reduction of
the kinesin binding-rates, compared to the experimentally observed
values (although this would be partly compensated by shorter run
lengths in bidirectional transport). With this caveat in mind, the
model results do indicate one \emph{possible} mechanism for
tau-induced transport disruption in neural diseases.  On one hand, the
robustness of the transport capacity for a wide range of tau-densities
indicates that the system is stable against small fluctuations of the
tau-concentration.  Bearing in mind that tau proteins stabilize the
network of microtubule filaments, it is expected that the transport
capacities of nerve cells are only marginally influenced by the
tau-concentration for non-pathological conditions. On the other hand,
for low motor densities, a more extreme overexpression of tau can
drastically reduce the transport properties of the cell and may
trigger a cascade of events leading to a loss of functionality---see,
e.g.,~\cite{Shemesh08} for alternative mechanisms of such cascades.

To conclude, we have presented a model approach to intracellular transport, which
supports the possible importance of tau proteins for neural diseases. In
future work the present approach should be extended, e.g., the structure of
the nerve cell should be described in more detail, in order to obtain a more
quantitative description of the relevant processes.

\vspace{\baselineskip}


\acknowledgments

We thank the DFG Research Training Group GRK 1276 for financial support.

\appendix

\section{Mean square sojourn time in unbound state}

Here we summarize the intermediate steps leading to the
result~\eqref{e:tu2} for the mean square sojourn time in the upper
lane.

Starting from~\eqref{e:tuN} one obtains (under approximation 1)
\begin{align}
\langle t_u(N)^2 \rangle_1 = & \langle T(N)^2 \rangle + \langle m + 1 \rangle
\langle T_\text{tr}^2  \rangle \notag \\ &+ \langle m(m + 1) \rangle \langle
T_\text{tr} \rangle^2 + \langle m \rangle \langle T_1(N)^2 \rangle \notag \\ &+
\langle m(m-1) \rangle \langle T_1(N) \rangle^2 \notag \\ &+ 2 \langle m + 1
\rangle \langle T(N) \rangle \langle T_\text{tr} \rangle \notag \\ &+ 2 \langle m 
\rangle \langle T(N) \rangle \langle T_1(N) \rangle \notag \\ &+ 2 \langle m(m+1)
\rangle \langle T_1(N) \rangle \langle T_\text{tr} \rangle.
\end{align}
In addition to quantities already introduced, we need the following averages
\begin{align}
\langle m^2 \rangle &= \frac{2-3A+A^2}{A^2} \\ 
\langle T_\text{tr}^2 \rangle &= \frac{2}{(2D+\omega_a)^2} \\
\langle T_l(N)^2 \rangle &= \frac{l(N-l)[N^2+1+l(N-l)]}{12 D^2} \\
\langle T(N)^2 \rangle &= \frac{N^4-1}{60D^2} 
\end{align}
(with the latter two obtained following~\cite{Weiss81}).
Finally, averaging over the distribution $P_0(N)$ of
line-segment lengths yields
\begin{align}
\langle t_u^2 \rangle_2 =& \sum_{N=1}^\infty P_0(N) \langle t_u(N)^2
\rangle_1 \\
=&  \{ \omega_a^2
  \rho_\tau(1+\rho_\tau)^2  +
  8D\omega_a\rho_\tau(1-\rho_\tau^2)
\\ &+4D^2(1-\rho_\tau)^2(1+4\rho_\tau+\rho_\tau^2)
  \}
\\
  &\times \{2(1-\rho_\tau)^4 D^2 \omega_a^2\}^{-1}.
\end{align}

The analogous calculation under approximation 2 gives instead
\begin{multline}
\langle t_u^2 \rangle_2 =  \{ \omega_a^2
  \rho_\tau(1+\rho_\tau)^2  +
  4D\omega_a\rho_\tau(1-\rho_\tau)
\\ +4D^2(1-\rho_\tau)^2
  \}
\\
  \times \{2(1-\rho_\tau)^4 D^2 \omega_a^2\}^{-1}.
\end{multline}

\bibliographystyle{apsrev}
\bibliography{/home/network/harris/allref}

\end{document}